\documentclass[a4paper,prb,reprint,showpacs,superscriptaddress]{revtex4-1}
\usepackage{graphicx,color}
\usepackage{amsmath,amssymb,bm}

\newcommand{\ket}[1]{\left|#1\right\rangle}
\newcommand{\bra}[1]{\left\langle#1\right|}
\newcommand{\ii}{\text{i}}

\begin{document}
\title{Dynamical quantum phase transitions in the quantum Potts chain}
\author{C. Karrasch}
\affiliation{Dahlem Center for Complex Quantum Systems and Fachbereich Physik, Freie Universit\"at Berlin, 14195 Berlin, Germany}
\author{D. Schuricht}
\affiliation{Institute for Theoretical Physics, Center for Extreme Matter and Emergent Phenomena, Utrecht University, Princetonplein 5, 3584 CE Utrecht, The Netherlands}

\begin{abstract}
We analyze the dynamics of the return amplitude following a sudden quench in the three-state quantum Potts chain. For quenches crossing the quantum critical point from the paramagnetic to the ferromagnetic phase, the corresponding rate function is non-analytic at critical times and behaves \emph{linearly} in their vicinity. In particular, we find no indication of a link between the time evolution close to the critical times and the scaling properties of the quantum critical point in the Potts chain. 
\end{abstract}

\maketitle


\section{Introduction}
Over the last decade, there has been a tremendous interest in the dynamics of quantum systems after the sudden change of one of its parameters (quantum quenches).\cite{CalabreseCardy06} One object that has been studied intensively is the return amplitude\cite{Silva08}
\begin{equation}
G(t)=\bra{\Psi_0}e^{-\ii Ht}\ket{\Psi_0}
\label{eq:overlapamplitude}
\end{equation}
which describes the overlap of the time-evolved state $\ket{\Psi(t)}=e^{-\ii Ht}\ket{\Psi_0}$ with the initial one $\ket{\Psi_0}$. In particular, Heyl et al.\cite{Heyl-13} investigated the formal similarity between the boundary partition function\cite{LeClair-95}
\begin{equation}
Z(z)=\bra{\Psi_0}e^{-zH}\ket{\Psi_0}
\label{eq:boundarypartitionfunction}
\end{equation}
and the return amplitude \eqref{eq:overlapamplitude} under the analytic continuation $z=\ii t$. For quenches in the transverse-field Ising chain, they showed that the lines of zeros of Eq.~\eqref{eq:boundarypartitionfunction} cross the imaginary axis at specific points $z=\ii t_n^\ast$ provided the quench is between different phases of the system. These crossings imply the existence of critical times $t_n^\ast$ at which the rate function 
\begin{equation}
l(t)=-\frac{1}{L}\ln\big|G(t)\big|^2,
\label{eq:ratefunction}
\end{equation}
with $L$ denoting the length of the system, shows a non-analytic behavior $l(t)\sim |t-t_n^\ast|$ for $t\to t_n^\ast$.\cite{Heyl15} This observation was coined\cite{Heyl-13} \emph{dynamical quantum phase transition} (DQPT). Various aspects of DQPTs were subsequently investigated in several other systems;\cite{KS13,Heyl14,Canovi-14,Hickey-14,AndraschkoSirker14,VajnaDora14,Kriel-14,VajnaDora15,SchmittKehrein15,Sharma-15,JamesKonik15,HuangBalatsky16,Divakaran-16,BudichHeyl16,AbelingKehrein16,Sharma-16,PS16,Zunkovic-16,Piroli-16} in particular, it was shown that non-analyticities in the rate function \eqref{eq:ratefunction} can also show up in the dynamics after quenches that do not cross a quantum phase transition.\cite{VajnaDora14,AndraschkoSirker14,Hickey-14} Significant progress has been made regarding the experimental observation of DQPTs in atomic\cite{Flaeschner-16} and ionic\cite{Jurcevic-16} systems in optical lattices.

Recently, Heyl\cite{Heyl15} employed renormalization-group (RG) arguments to propose a relation between DQPTs and unstable fixed points of the RG flow of the underlying model in thermal equilibrium. For example, he argued that the rate function \eqref{eq:ratefunction} in the prototypical transverse-field Ising chain close to the critical times is given by
\begin{equation}
\label{eq:heylrg}
l(t)\sim |t-t_n^\ast|^{d/y} 
\end{equation}
with $d=1$ the dimension of the system and $y=1$ the RG eigenvalue at the unstable fixed point. This linear behavior is consistent with earlier results for the transverse-field Ising chain\cite{Silva08,Heyl-13} as well as for the axial next-nearest-neighbour Ising model.\cite{KS13}

It is an interesting question whether systems exist that exhibit DQPTs with \emph{non-linear} scaling behavior close to the critical times. In light of the conjectured relation between DQPTs and the universal properties in equilibrium, a natural candidate is a model featuring a quantum critical point with a non-trivial RG eigenvalue $y$. This motivates us to investigate the dynamics of the three-state quantum Potts chain,\cite{GehlenRittenberg86,Mong-14} which possesses a paramagnetic (PM) and a ferromagnetic (FM) phase separated by a quantum critical point with $y=6/5$.\cite{Dotsenko84,DiFrancescoMathieuSenechal97,Mussardo10} For quenches from the PM to the FM phase, we find that the rate function shows non-analyticities at critical times $t^\ast$ but behaves linearly in their close vicinity, $l(t)\sim |t-t^\ast|$. This result is supported both by an exact analytical solution that we obtain for special quench parameters as well as by numerical matrix product state simulations for general situations. In other words, there seems to be \emph{no relation} between the time evolution of the rate function $l(t)$ close to the critical times $t^\ast$ and the scaling properties of the quantum critical point in the three-state quantum Potts chain. For opposite quenches from the FM to PM phase, the behavior is more complicated. If one starts from a fully-polarized FM state and quenches to a system with infinite transverse field, no non-analyticities appear; if the quench ends at a finite transverse field, a time scale exists after which non-analytic behavior develops. 

This article is organized as follows: In Sec.~\ref{sec:model}, we first review the some facts about the three-state quantum Potts chain. In Sec.~\ref{sec:PMFM}, we present our results for quenches from the PM phase to the FM phase, which contain both exact analytical results for a special limit as well as numerical simulations obtained using a standard time-dependent density-matrix renormalization-group (DMRG) algorithm.\cite{Vidal04,tdmrg2,Daley-04,Schmitteckert04} In Sec.~\ref{sec:discussion}, we provide a discussion of our results in light of the previous RG analysis.\cite{Heyl15} In Sec.~\ref{sec:FMPM}, the opposite quench from the FM to the PM is studied, before we conclude in Sec.~\ref{sec:conclusion}. 

\section{Three-state quantum Potts chain}\label{sec:model}
In this work, we consider the three-state quantum Potts chain defined by the Hamiltonian\cite{GehlenRittenberg86,Mong-14}
\begin{equation}
H=-J\sum_i\bigl(\sigma_i^\dagger\sigma_{i+1}+\sigma_{i+1}^\dagger\sigma_i\bigr)-f\sum_i\bigl(\tau_i^\dagger+\tau_i\bigr),
\label{eq:model}
\end{equation}
where the operators $\sigma_i$ and $\tau_i$ act on the three states of the local Hilbert space at site $i$, which we label by $\ket{A}_i$, $\ket{B}_i$, and $\ket{C}_i$, respectively. Their explicit matrix representation is given by
\begin{equation}
\sigma=\left(\begin{array}{ccc}1 & 0 & 0\\ 0 & \omega & 0\\ 0 & 0 & \omega^2\end{array}\right),\quad
\tau=\left(\begin{array}{ccc}0 & 0 & 1\\ 1 & 0 & 0\\ 0 & 1 & 0\end{array}\right),
\end{equation}
with $\omega=\exp(2\pi\ii/3)$. We assume $J,f\ge 0$. For $f<J$, the model possesses a FM phase with a three-fold degenerate ground state that spontaneously breaks the S$_3$ symmetry. For $f>J$, the model is in a PM regime with a unique ground state. The two phases are separated by a quantum critical point at $f=f_\text{c}=J$ which is described by the minimal conformal field theory\cite{Dotsenko84,DiFrancescoMathieuSenechal97,Mussardo10} with central charge $c=4/5$ and $\nu=1/y=5/6$ the critical exponent of the correlation length; i.e., close to the critical point, the correlation length diverges as $\xi\sim |f-f_\text{c}|^{-\nu}$. 

We now study the time evolution of the return amplitude \eqref{eq:overlapamplitude} after sudden quenches between the two phases; we begin with quenches from the PM to the FM side. 

\section{Quench PM to FM}\label{sec:PMFM}
In this section, we investigate DQPTs after quenches from the PM to the FM phase. The quench protocol is implemented by suddenly switching the transverse field $f$ from its initial value $f=f_0$ to its final value $f=f_1$ while keeping the exchange interaction $J$ constant. 

\begin{figure}[t]
\includegraphics[width=0.95\linewidth,clip]{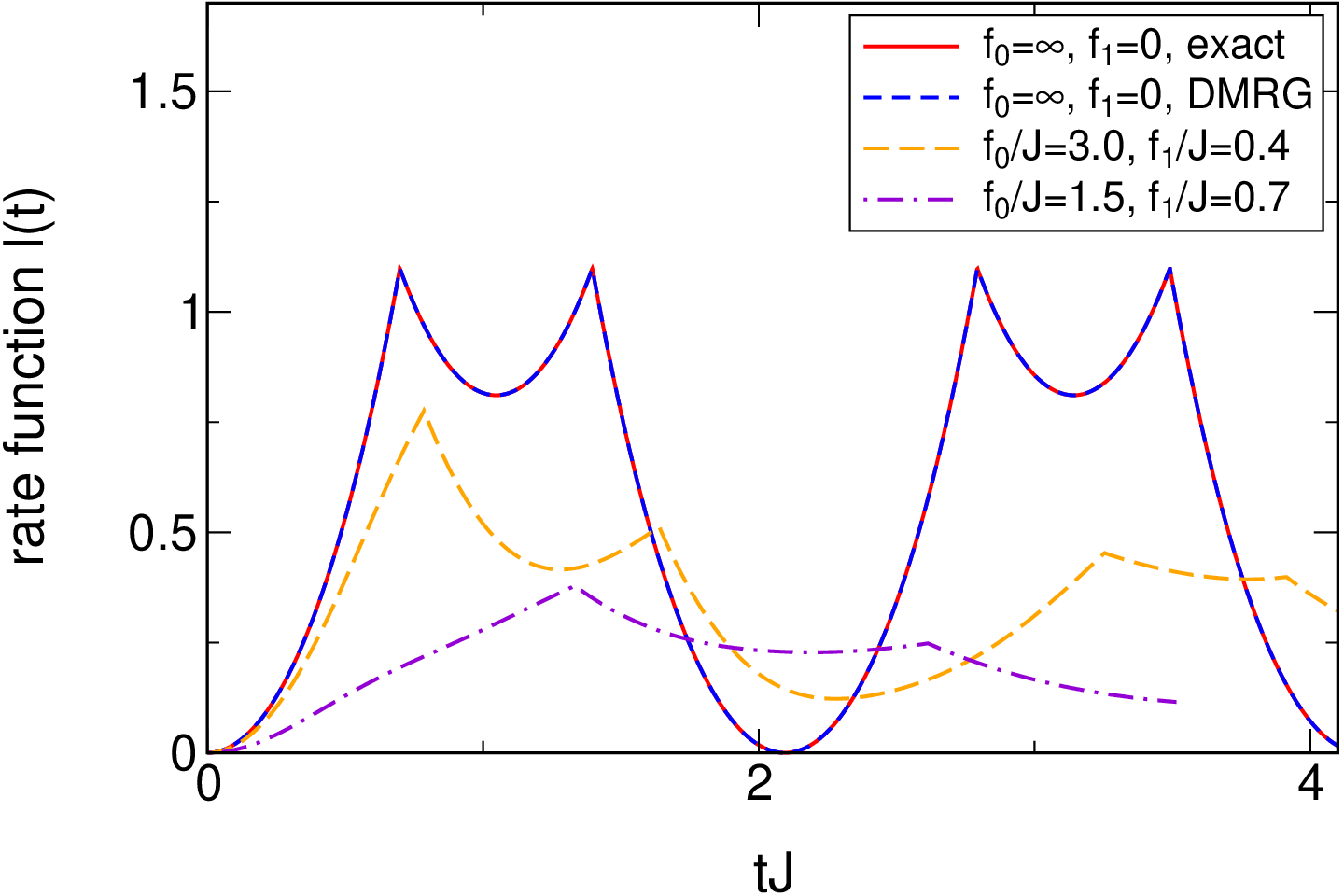}
\caption{(Color online) Rate function $l(t)$ defined in Eq.~\eqref{eq:ratefunction} for a quench from the PM to the FM phase in the three-state quantum Potts chain. $l(t)$ characterizes the overlap with the initial state in the thermodynamic limit. The DMRG data agree well with the analytic result \eqref{eq:analyticPMFM} obtained for the trivial quench $f_0=\infty$ to $f_1=0$. The rate function shows non-analytic behavior (kinks) at the critical times $t^*$.}
\label{fig:figure1}
\end{figure}
\begin{figure}[t]
\includegraphics[width=0.95\linewidth,clip]{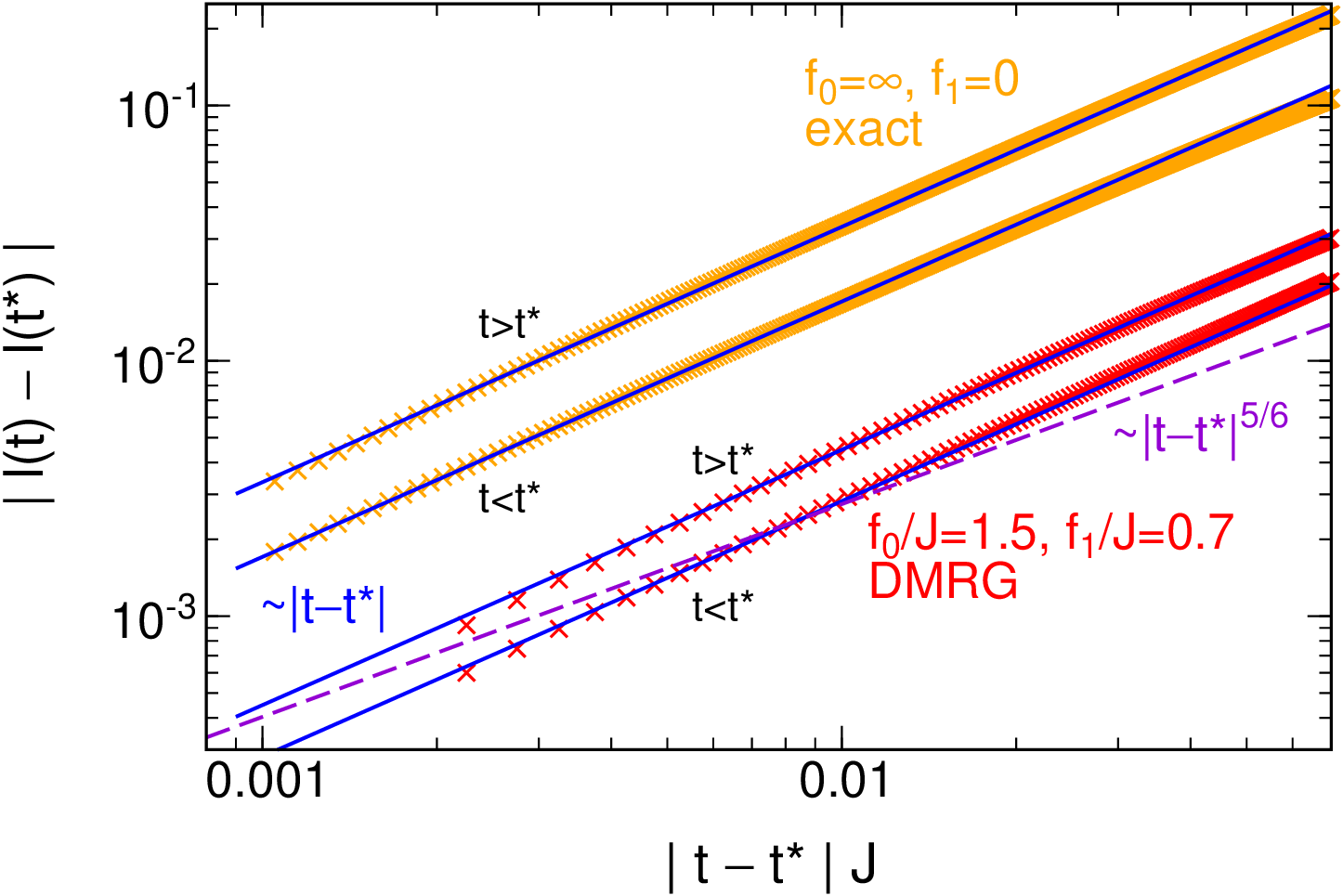}
\caption{(Color online) The rate function $l(t)$ of Fig.~\ref{fig:figure1} close to the first critical time $t^\ast$. The behavior is always linear, $l(t)\propto|t-t^\ast|$. For comparison, we also show the naive expectation predicted by Eq.~\eqref{eq:heylrg} with $d/y=5/6$.}
\label{fig:figure2}
\end{figure}

First, we consider a special limit in which the rate function \eqref{eq:ratefunction} can be obtained analytically: We start from the perfect PM state ($f_0=\infty$) and quench to the classical FM chain ($f_1=0$). The initial state is given by
\begin{equation}
\ket{\Psi_0}=\frac{1}{3^{L/2}}\prod_i\bigl(\ket{A}_i+\ket{B}_i+\ket{C}_i\bigr).
\end{equation}
Since the final Hamiltonian is purely classical, the return amplitude \eqref{eq:overlapamplitude} takes the simple form
\begin{equation}
G(t)=\text{tr}\,M^L,\quad M=\frac{1}{3}\left(\begin{array}{ccc}e^{2\ii Jt}&e^{-\ii Jt}&e^{-\ii Jt}\\ e^{-\ii Jt}&e^{2\ii Jt}&e^{-\ii Jt}\\ e^{-\ii Jt}&e^{-\ii Jt}&e^{2\ii Jt}\end{array}\right),
\end{equation}
where periodic boundary conditions on a chain with $L$ lattice sites have been imposed. The eigenvalues of the auxiliary matrix $M$ are given by 
\begin{equation}
\Lambda_1=\frac{e^{-\ii Jt}}{3}\bigl(e^{3\ii Jt}+2\bigr),\quad\Lambda_2=\Lambda_3=\frac{e^{-\ii Jt}}{3}\bigl(e^{3\ii Jt}-1\bigr).
\end{equation}
Thus, we obtain
\begin{equation}
G(t)=\Lambda_1(t)^L+2\Lambda_2(t)^L,
\end{equation}
which yields the rate function 
\begin{equation}
\begin{split}
l(t)=&-\frac{1}{L}\ln\Big|\bigl(9\cos^2\hat{t}+\sin^2\hat{t}\bigr)^L+4^{L+1}\sin^{2L}\hat{t}\\*
&\qquad\qquad +2(2\ii)^L\bigl(3\cos\hat{t}+\ii\sin\hat{t}\bigr)^L\sin^L\hat{t}\\*
&\qquad\qquad +2(2\ii)^L\bigl(-3\cos\hat{t}+\ii\sin\hat{t}\bigr)^L\sin^L\hat{t}\,\Big|\\*
&\qquad+2\ln 3,
\end{split}
\label{eq:analyticPMFM}
\end{equation}
where $\hat{t}=3Jt/2$. The rate function \eqref{eq:analyticPMFM} is periodic, $l(t)=l(t+2\pi/(3J))$, satisfies $l(0)=0$, and shows non-analytic behavior at the critical times $Jt^\ast=2\pi/9+2\pi n/3$ as well as $Jt^\ast=4\pi/9+2\pi n/3$ with $n\in\mathbb{N}_0$. This is illustrated in Fig.~\ref{fig:figure1}. The critical times can be identified as the times at which the eigenvalues of the auxiliary matrix $M$ have equal modulus, $|\Lambda_1(t^\ast)|=|\Lambda_2(t^\ast)|$. Close to $t^\ast$, the rate function behaves linearly
\begin{equation}
l(t)\propto |t-t^\ast|,
\label{eq:linearkink}
\end{equation}
which is shown explicitly in Fig.~\ref{fig:figure2}.

For general quench parameters, the time evolution of the return amplitude can be computed using a standard time-dependent DMRG framework.\cite{Vidal04,tdmrg2,Daley-04,Schmitteckert04} We employ an 
infinite-system algorithm that is set up directly in the thermodynamic limit. The discarded weight is kept constant during the real time evolution, which leads to a dynamic increase of the bond dimension. We performed every calculation using three different values of the discarded weight in order to ensure convergence. Further details of the numerical implementation can be found in  Ref.~\onlinecite{KS13}.

Examples for the time evolution of the rate function for general quench parameters are shown in Fig.~\ref{fig:figure1}. We clearly observe non-analytic behavior at critical times $t^\ast$ for arbitrary $f_0$ and $f_1$. As shown in Fig.~\ref{fig:figure2}, the time evolution close to these critical times is again linear (small deviations are not related to the error of the raw DMRG data but most likely due to the fact that the critical time is determined numerically with an error equal to the finite resolution $\Delta t=0.0005$ of the time axis). The rate function does not show any sign of the non-trivial exponent $\nu=1/y=5/6$ governing the quantum critical point of the three-state Potts model. Thus, there seems to be \emph{no relation} between the dynamics of $l(t)$ close to $t^\ast$ and the properties of the quantum critical point in the three-state quantum Potts chain. 

\section{Discussion}\label{sec:discussion}
Our analytical and numerical calculations show that the return amplitude for quenches in the Potts model behaves linearly close to the critical times. In contrast, Heyl's interesting conjecture of Eq.~\eqref{eq:heylrg}, which is based on an RG treatment for the Ising model,\cite{Heyl15} naively suggests a power law with a non-trivial exponent of $5/6$. We now shed light on the origin of this putative discrepancy.

We first briefly review the RG analysis of Ref.~\onlinecite{Heyl15} for the transverse-field Ising chain governed by $H_\text{TFI}=-J\sum_{i}\sigma_i^z\sigma_{i+1}^z-g\sum_i\sigma_i^x$. The starting point is the observation that for a quench from a trivial PM state ($g\to\infty$) to the classical FM system ($g=0$), the return amplitude $G(t)$ is formally identical to the partition function of the classical model at a complex temperature $T=1/K$, i.e., $G(t)=Z(K)$ where $Z(K)=\text{tr}(e^{-KH_\text{TFI}})$ and $K=\ii Jt$. A standard block decimation\cite{Cardy96} is then used to eliminate every second site; the partition function remains unchanged (up to a multiplicative constant) provided that the new coupling constant $K'$ is related to the original one via $\tanh K'=\tanh^2 K$. This RG equation has two fixed points at $K=0$ and $K=\infty$ which correspond to infinite and zero temperatures, respectively. The relation to DQPTs is now made\cite{Heyl15} by the observation that the critical times $t_n^\ast$ map onto the $K=\infty$ fixed point after two RG steps. By analyzing the scaling behavior around $K=\infty$, Heyl finally obtains the power-law dependence of Eq.~(\ref{eq:heylrg}).

We stress that the relevant fixed point in Heyl's line of argument is the unstable, zero-temperature fixed point $K=1/T=\infty$ of the classical Ising chain with $g=0$, not the quantum critical point of the quantum Ising chain (which is located at $T=0$ and $g=1$). Thus, to the best of our understanding, the RG analysis for the Ising chain does not provide a relation between the dynamics of the rate function and the quantum phase transition. Furthermore, we note that in thermal equilibrium, the behavior around the zero-temperature fixed point $K=\infty$ is exponential\cite{Mussardo10} in the temperature (and thus in $K$); hence, the extraction of the power-law  behavior in Eq.~(\ref{eq:heylrg}) seems mathematically quite subtle. 
\begin{figure*}[t]
\includegraphics[width=0.48\linewidth,clip]{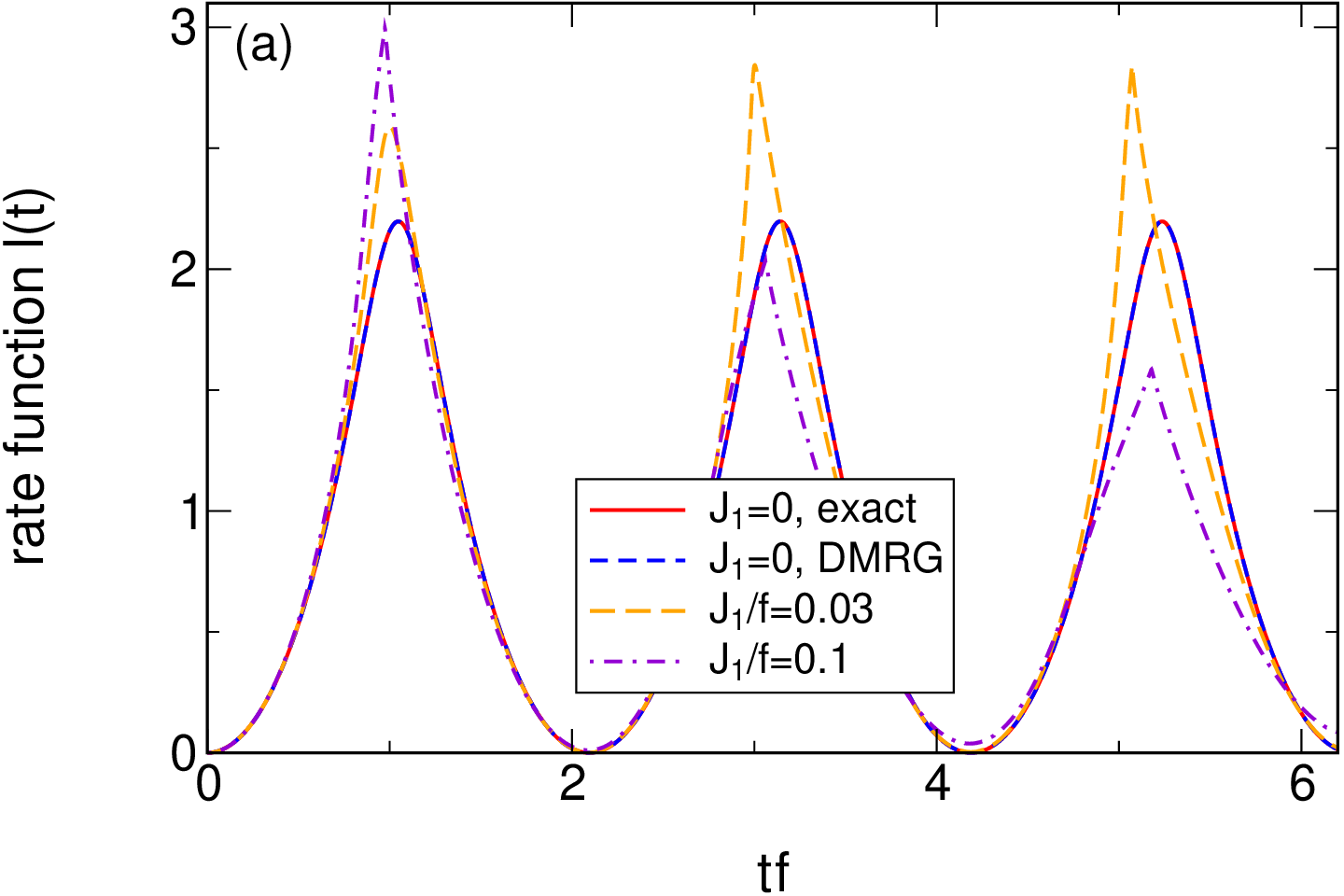}
\hspace*{0.02\linewidth}
\includegraphics[width=0.48\linewidth,clip]{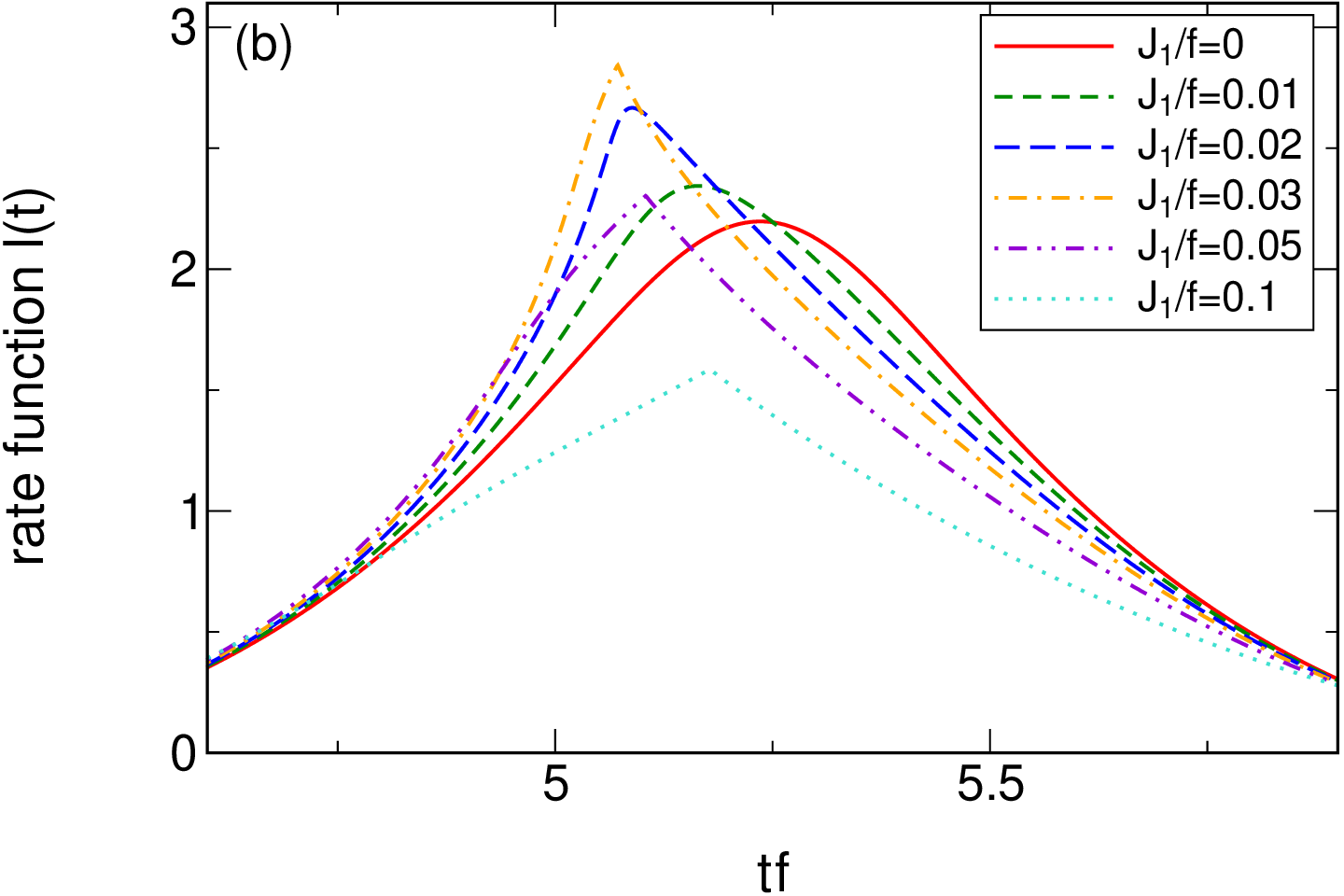}
\caption{(Color online) Rate function $l(t)$ for a quench starting from a fully-polarized FM state to the Potts chain with final parameters $J_1$ and $f$ (see text). For the trivial quench to $J_1=0$, $l(t)$ is analytic at all times [see also Eq.~\eqref{eq:ratefunctionFMPM}]. For finite $J_1$, non-analyticities (kinks) develop for large enough times.}
\label{fig:figure3}
\end{figure*}

To sum up, Heyl's RG treatment (for the Ising chain) does not involve the quantum critical point, thus there is in fact no discrepancy between our results for the Potts chain and his approach.

\section{Quench FM to PM}\label{sec:FMPM}
In this section, we study DQPTs after quenches from the FM to the PM phase. The quench protocol is implemented by suddenly switching the exchange interaction $J$ from the initial value $J=J_0$ to its final value $J=J_1$ while keeping the transverse field $f$ constant. 

As initial state we use the fully-polarized pure state 
\begin{equation}
\ket{\Psi_0}=\prod_i\ket{A}_i,
\end{equation}
which is obtained as one of the ground states in the limit $J_0\to\infty$. We quench to the PM phase of the Potts chain ($J_1<f$). In the case of the trivial quench to $J_1=0$, the rate function can again be calculated analytically with the result
\begin{equation}
l(t)=2\ln 3-\ln\big|5+4\cos(3Jt)\big|.
\label{eq:ratefunctionFMPM}
\end{equation}
We stress that $l(t)$ is completely smooth at all times even though the quench crosses the quantum critical point. Similar behavior has previously been observed in the XY and XXZ spin chains.\cite{VajnaDora14,AndraschkoSirker14} For finite $J_1$, we calculate the rate function via the time-dependent DMRG; the results are shown in Fig.~\ref{fig:figure3}. Interestingly, for sufficiently small $J_1$, the first few maxima in the rate function are still smooth while kinks develop at later times. The relation between the value of $J_1$ and the time $t_\text{c}$ after which non-analyticities are observed is approximately given by $t_\text{c}\sim 1/J_1$. The behavior close to the critical times again seems to be linear (which we have checked for a limited set of parameters).

\section{Conclusion}\label{sec:conclusion}
We have calculated the return amplitude after sudden quenches in the three-state quantum Potts chain. This model possesses a FM and a PM phase separated by a quantum phase transition with non-trivial (i.e., non-integer) critical exponents. For quenches from the PM to the FM regime, the rate function $l(t)$ possesses non-analytic kinks at critical times $t^\ast$, close to which the behavior is linear, $l(t)\sim|t-t^\ast|$. In other words, the non-analyticities do not show any signature of the non-trivial exponents of the quantum critical point. For the quench from the FM to the PM regime, there exists a critical time $t_\text{c}$ until which the rate function remains smooth, while for later times the kinks reappear.

\acknowledgments
We thank Paul Fendley, Lars Fritz, Markus Heyl and Volker Meden for very useful comments and discussions. CK is supported by the DFG via the Emmy-Noether program under KA 3360/2-1. DS is member of the D-ITP consortium, a program of the Netherlands Organisation for Scientific Research (NWO) that is funded by the Dutch Ministry of Education, Culture and Science (OCW). DS was supported by the Foundation for Fundamental Research on Matter (FOM), which is part of the Netherlands Organisation for Scientific Research (NWO), under 14PR3168.



\begin{thebibliography}{10}

\bibitem{CalabreseCardy06}
P. Calabrese and J. Cardy, Phys. Rev. Lett. {\bf 96},  136801  (2006).

\bibitem{Silva08}
A. Silva, Phys. Rev. Lett. {\bf 101},  120603  (2008).

\bibitem{Heyl-13}
M. Heyl, A. Polkovnikov, and S. Kehrein, Phys. Rev. Lett. {\bf 110},  135704
  (2013).

\bibitem{LeClair-95}
A. LeClair, G. Mussardo, H. Saleur, and S. Skorik, Nucl. Phys. B {\bf 453},
  581  (1995).

\bibitem{Heyl15}
M. Heyl, Phys. Rev. Lett. {\bf 115},  140602  (2015).

\bibitem{KS13}
C. Karrasch and D. Schuricht, Phys. Rev. B {\bf 87},  195104  (2013).

\bibitem{Heyl14}
M. Heyl, Phys. Rev. Lett. {\bf 113},  205701  (2014).

\bibitem{Canovi-14}
E. Canovi, P. Werner, and M. Eckstein, Phys. Rev. Lett. {\bf 113},  265702
  (2014).

\bibitem{Hickey-14}
J.~M. Hickey, S. Genway, and J.~P. Garrahan, Phys. Rev. B {\bf 89},  054301
  (2014).

\bibitem{AndraschkoSirker14}
F. Andraschko and J. Sirker, Phys. Rev. B {\bf 89},  125120  (2014).

\bibitem{VajnaDora14}
S. Vajna and B. D\'ora, Phys. Rev. B {\bf 89},  161105  (2014).

\bibitem{Kriel-14}
J.~N. Kriel, C. Karrasch, and S. Kehrein, Phys. Rev. B {\bf 90},  125106
  (2014).

\bibitem{VajnaDora15}
S. Vajna and B. D\'ora, Phys. Rev. B {\bf 91},  155127  (2015).

\bibitem{SchmittKehrein15}
M. Schmitt and S. Kehrein, Phys. Rev. B {\bf 92},  075114  (2015).

\bibitem{Sharma-15}
S. Sharma, S. Suzuki, and A. Dutta, Phys. Rev. B {\bf 92},  104306  (2015).

\bibitem{JamesKonik15}
A.~J.~A. James and R.~M. Konik, Phys. Rev. B {\bf 92},  161111  (2015).

\bibitem{HuangBalatsky16}
Z. Huang and A.~V. Balatsky, Phys. Rev. Lett. {\bf 117},  086802  (2016).

\bibitem{Divakaran-16}
U. Divakaran, S. Sharma, and A. Dutta, Phys. Rev. E {\bf 93},  052133  (2016).

\bibitem{BudichHeyl16}
J.~C. Budich and M. Heyl, Phys. Rev. B {\bf 93},  085416  (2016).

\bibitem{AbelingKehrein16}
N.~O. Abeling and S. Kehrein, Phys. Rev. B {\bf 93},  104302  (2016).

\bibitem{Sharma-16}
S. Sharma, U. Divakaran, A. Polkovnikov, and A. Dutta, Phys. Rev. B {\bf 93},
  144306  (2016).

\bibitem{PS16}
T. Pu\v{s}karov and D. Schuricht, SciPost Phys. {\bf 1},  003  (2016).

\bibitem{Zunkovic-16}
B. Zunkovic, M. Heyl, M. Knap, and A. Silva, arXiv:1609.08482.

\bibitem{Piroli-16}
L. Piroli, B. Pozsgay, and E. Vernier, arXiv:1611.06126.

\bibitem{Flaeschner-16}
N. Fl\"aschner, D. Vogel, M. Tarnowski, B.~S. Rem, D.-S. L\"uhmann, M. Heyl,
  J.~C. Budich, L. Mathey, K. Sengstock, and C. Weitenberg, arXiv:1608.05616.

\bibitem{Jurcevic-16}
P. Jurcevic, H. Shen, P. Hauke, C. Maier, T. Brydges, C. Hempel, B. P. Lanyon, M. Heyl, R. Blatt, and C. F. Roos, arXiv:1612.06902.

\bibitem{GehlenRittenberg86}
G. von Gehlen and V. Rittenberg, J. Phys. A {\bf 19},  L625  (1986).

\bibitem{Mong-14}
R.~S.~K. Mong {\it et~al.}, Phys. Rev. X {\bf 4},  011036  (2014).

\bibitem{Dotsenko84}
V.~S. Dotsenko, Nucl. Phys. B {\bf 235},  54  (1984).

\bibitem{DiFrancescoMathieuSenechal97}
P. Di~Francesco, P. Mathieu, and D. S\'{e}n\'{e}chal, {\em {C}onformal {F}ield
  {T}heory} (Springer, New York, 1997).

\bibitem{Mussardo10}
G. Mussardo, {\em {S}tatistical {F}ield {T}heory} (Oxford University Press,
  Oxford, 2010).

\bibitem{Vidal04}
G. Vidal, Phys. Rev. Lett. {\bf 93},  040502  (2004).

\bibitem{tdmrg2}
   S.~R.~White and A.~E.~Feiguin, Phys. Rev. Lett. {\bf 93}, 076401 (2004).

\bibitem{Daley-04}
A.~J. Daley, C. Kollath, U. Schollw\"ock, and G. Vidal, J. Stat. Mech. P04005 (2004).

\bibitem{Schmitteckert04}
P. Schmitteckert, Phys. Rev. B {\bf 70},  121302(R)  (2004).

\bibitem{Cardy96}
J. Cardy, {\em {S}caling and {R}enormalization in {S}tatistical {P}hysics}
  (Cambridge University Press, Cambridge, 1996).

\end{thebibliography}
\end{document}